\documentclass[conference]{IEEEtran}
\IEEEoverridecommandlockouts
\usepackage{cite}
\usepackage{amsmath,amssymb,amsfonts}
\usepackage{graphicx}
\usepackage{xcolor}
\usepackage{comment, lipsum}
\usepackage{amsmath,amssymb} % define this before the line numbering.
\usepackage{tikz}
\usepackage{color}
\usepackage{multirow, multicol}
\usepackage{graphicx, caption, subcaption}
\usepackage{newtxtext,newtxmath}
\usepackage{subcaption}
\usepackage{newtxmath}
\usepackage{algorithm}% http://ctan.org/pkg/algorithm
\usepackage{algpseudocode}% http://ctan.org/pkg/algorithmicx
\usepackage{varwidth}
\usepackage{tabulary,booktabs}

\begin{document}

\title{Memristive Computing for Efficient Inference on Resource Constrained Devices
}

\author{
\IEEEauthorblockN{Venkatesh Rammamoorthy}
\IEEEauthorblockA{
\textit{University of Central Florida} \\
Florida, USA}
\and
\IEEEauthorblockN{Geng Zhao}
\IEEEauthorblockA{
\textit{University of Central Florida} \\
Florida, USA}
\and
\IEEEauthorblockN{Bharathi Reddy}
\IEEEauthorblockA{
\textit{Sundrape Inc.}\\
Bangalore, India}
\and
\IEEEauthorblockN{Ming-Yang Lin}
\IEEEauthorblockA{
\textit{Nanjing University} \\
Nanjing, China}
}

\maketitle

\begin{abstract}
The advent of deep learning has resulted in a number of applications which have transformed the landscape of the research area in which it has been applied. However, with an increase in popularity, the complexity of classical deep neural networks has increased over the years. As a result, this has leads to considerable problems during deployment on devices with space and time constraints. In this work, we perform a review of the present advancements in non-volatile memory and how the use of resistive RAM memory, particularly memristors, can help to progress the state of research in deep learning. In other words, we wish to present an ideology that advances in the field of memristive technology can greatly influence and impact deep learning inference on edge devices.
\end{abstract}

\section{Introduction}

\begin{figure*}
    \centering
    \includegraphics{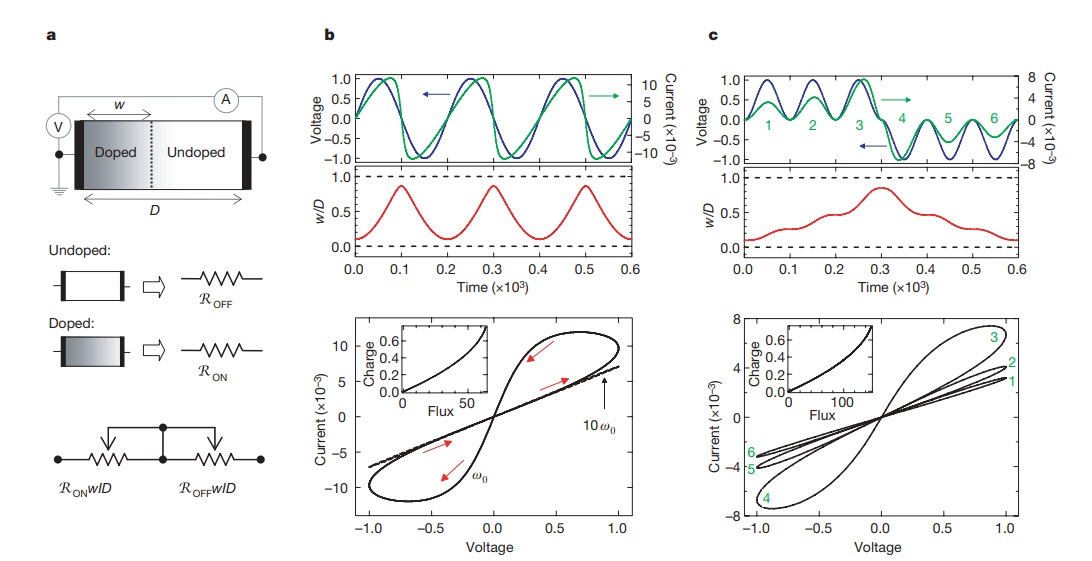}
    \caption{The IV characteristic of a memristor - shown as the missing element of the circuit. Additionally, the memristor has also been shown in terms of a programmable resistance based on the amount of doping in the memristor.}
    \label{fig:memristors}
\end{figure*}

Memristors were first proposed by Leon Chua \cite{chua1971memristor} as the missing element of the circuit, which can be used to change its resistance based on the amount of current passing through it. This spurred an interesting line of research on using the concept of spike-timing-dependent-plasticity (STDP) to program the resistance of the memristor \cite{serrano2013stdp, acciarito2016vlsi, panwar2017arbitrary}. This enabled several methods to use the properties of a memristor to emulate an artificial biological synapse. Fig. \ref{fig:memristors} shows the resistive memory model for a memristor, in which the resistance of the model is coupled with the voltage across it. Interestingly, there have been previous papers which have supported the positive impact of advancement in resistive memory on deep learning, particularly computer vision \cite{subramaniam2017neuromorphic}, further proposing the impact in medical and sport sciences as well \cite{subramaniam2019spectral,  balakrishnan2013detecting,icaart}.

Interesting, a few papers have employed memristors to realize highly sparse and efficient convolutional neural networks which can easily be deployed on devices with significant time and space constraints. Some approaches have designed a memristor-based sparse compact convolutional neural network (MSCCNN) to reduce the number of memristors \cite{8798700}. An average pooling and 1 × 1 convolutional layer are used in order to replace fully connected layers. Meanwhile, depthwise separation convolution is utilized to replace traditional convolution to further reduce the number of parameters. Furthermore, a network pruning method is adopted to remove the redundant memristor crossbars for depthwise separation convolutional layers. Therefore, a more compact network structure is obtained while the recognition accuracy remaining unchanged. Another approach uses memristors to perform pruning as well as quantization in deep neural networks by incorporating alternating direction method of multipliers (ADMM) while training the deep neural network.  

Interesting, a few papers have employed memristors to realize highly sparse and efficient convolutional neural networks which can easily be deployed on devices with significant time and space constraints. this paper aims to design a memristor-based sparse compact convolutional neural network (MSCCNN) to reduce the number of memristors \cite{8798700}. An average pooling and 1 × 1 convolutional layer are used in order to replace fully connected layers. Meanwhile, depthwise separation convolution is utilized to replace traditional convolution to further reduce the number of parameters. Furthermore, a network pruning method is adopted to remove the redundant memristor crossbars for depthwise separation convolutional layers. Therefore, a more compact network structure is obtained while the recognition accuracy remaining unchanged. 

Another approach uses memristors to perform pruning as well as quantization in deep neural networks by incorporating alternating direction method of multipliers (ADMM) while training the deep neural network \cite{8824944}. We consider hardware constraints such as crossbar blocks pruning, conductance range, and mismatch between weight value and real devices, to achieve high accuracy and low power and small area footprint. 

The framework primarily consists of three main steps:  
\begin{itemize}
    \item Memristor-based ADMM regularized optimization 
    \item Masked mapping
    \item Retraining
\end{itemize}

The above method was able to achieve an impressive compression rate of over 20$\times$ with respect to the weight compression rate and a compression of 24$\times$ with regard to the weight quantization of the deep learning models. Table 1 shows the results of the model on CIFAR-10 and MNIST datasets for deep learning models such as LeNet-5, ResNet-18 and VGG-16. While several papers have explored more complicated methods of pruning such as pruning during training \cite{lecun1990optimal, hassibi1993optimal}, pruning after training (in which the model is pruned after it is completely trained, followed by a fine-tuning procedure to ensure convergence of the pruned model), pruning before training \cite{lee2018snip, subramaniam2020n2nskip}, hardware-based methods have gone for more conventional methods of pruning. These methods have mostly adhered to iterative pruning strategies in which the model is either trained and pruned multiple times, or the model is pruned after the corresponding reference model achieves convergence on the dataset.
For instance, in \cite{acciarito2016vlsi}, an average pooling and 1 × 1 convolutional layer are used in order to replace fully connected layers. Meanwhile, depthwise separation convolution is utilized to replace traditional convolution to further reduce the number of parameters. Furthermore, a network pruning method is adopted to remove the redundant memristor crossbars for depthwise separation convolutional layers. Therefore, a more compact network structure is obtained while the recognition accuracy remaining unchanged. 

Another related field in which memristors have found a lot of impact has been neuromorphics \cite{memNeuro1, memNeuro2, memNeuro3, memNeuro4, memNeuro5}. Computation using memristive systems provide an interesting and compelling solution to increase the overall efficiency of the neuromorphic system. While some papers have identified some gaps in the behavior-level dynamics of memristive computing systems during simulation and proposed solutions, other have used different fabrication techniques to apply memristors for more machine learning based applications. Some of these applications include implementing well-known machine learning algorithms such as a multilayer perceptron network, K Nearest Neighbors and Independent Component Analysis \cite{raj2019programming, boppidi2020implementation}.

Memristor-based computation provides a promising
solution to boost the power efficiency of the neuromorphic computing system.

\begin{figure*}
    \centering
    \includegraphics[scale=0.8]{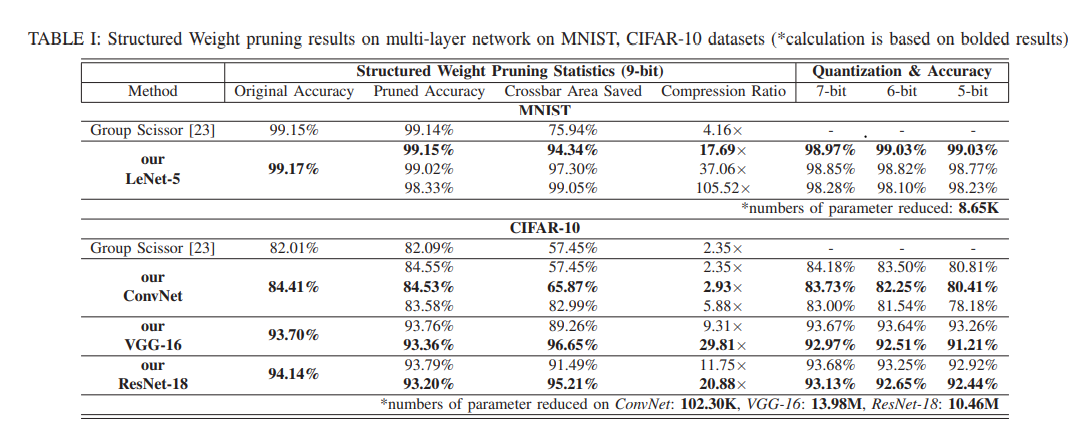}
    \label{fig:memristor}
\end{figure*}

\section{Conclusion}
In this study, we examine the state of research in resistive RAM technology, particularly in the field of memristors. We found that a majority of approaches have used the analog properties of memristors to emulate the learning capabilities of a biological synapse. A number of studies have shown that Memristor-based matrix-vector multiplication platform provides fast computation, high accuracy and low design cost.
However, as the development of memristor technology is still
maturing, device defects and fabrication yield may be a significant concern. Specifically, the single-bit failure (SBF)
denotes a device that freezes in a high conductance state
(“stuck-on”) or a low conductance state (“stuck-off”). Although neural networks usually can tolerate a certain number of imperfect synaptic weights, high SBF rate degrades
the computation accuracy significantly. For example, we
tested a feed-forward neural network for MNIST database:
as the SBF rate increases to 20\%, the average recognition
accuracy rapidly dropped from 92.64\% to 39.4\%, which is
far below an acceptable range. Redundancy schemes have
been widely adopted in memory designs. But it is not
efficient for the memristor-based analog computations with
high precision requirement.

\bibliography{aaai1}

\begin{thebibliography}{10}
\providecommand{\url}[1]{#1}
\csname url@samestyle\endcsname
\providecommand{\newblock}{\relax}
\providecommand{\bibinfo}[2]{#2}
\providecommand{\BIBentrySTDinterwordspacing}{\spaceskip=0pt\relax}
\providecommand{\BIBentryALTinterwordstretchfactor}{4}
\providecommand{\BIBentryALTinterwordspacing}{\spaceskip=\fontdimen2\font plus
\BIBentryALTinterwordstretchfactor\fontdimen3\font minus
  \fontdimen4\font\relax}
\providecommand{\BIBforeignlanguage}[2]{{%
\expandafter\ifx\csname l@#1\endcsname\relax
\typeout{** WARNING: IEEEtran.bst: No hyphenation pattern has been}%
\typeout{** loaded for the language `#1'. Using the pattern for}%
\typeout{** the default language instead.}%
\else
\language=\csname l@#1\endcsname
\fi
#2}}
\providecommand{\BIBdecl}{\relax}
\BIBdecl

\bibitem{chua1971memristor}
L.~Chua, ``Memristor-the missing circuit element,'' \emph{IEEE Transactions on
  circuit theory}, vol.~18, no.~5, pp. 507--519, 1971.

\bibitem{serrano2013stdp}
T.~Serrano-Gotarredona, T.~Masquelier, T.~Prodromakis, G.~Indiveri, and
  B.~Linares-Barranco, ``Stdp and stdp variations with memristors for spiking
  neuromorphic learning systems,'' \emph{Frontiers in neuroscience}, vol.~7,
  p.~2, 2013.

\bibitem{acciarito2016vlsi}
S.~Acciarito, A.~Cristini, L.~Di~Nunzio, G.~M. Khanal, and G.~Susi, ``An a vlsi
  driving circuit for memristor-based stdp,'' in \emph{2016 12th Conference on
  Ph. D. Research in Microelectronics and Electronics (PRIME)}.\hskip 1em plus
  0.5em minus 0.4em\relax IEEE, 2016, pp. 1--4.

\bibitem{panwar2017arbitrary}
N.~Panwar, B.~Rajendran, and U.~Ganguly, ``Arbitrary spike time dependent
  plasticity (stdp) in memristor by analog waveform engineering,'' \emph{IEEE
  Electron Device Letters}, vol.~38, no.~6, pp. 740--743, 2017.

\bibitem{subramaniam2017neuromorphic}
A.~Subramaniam, ``A neuromorphic approach to image processing and machine
  vision,'' in \emph{2017 Fourth International Conference on Image Information
  Processing (ICIIP)}.\hskip 1em plus 0.5em minus 0.4em\relax IEEE, 2017, pp.
  1--6.

\bibitem{subramaniam2019spectral}
A.~Subramaniam and K.~Rajitha, ``Spectral reflectance based heart rate
  measurement from facial video,'' in \emph{2019 IEEE International Conference
  on Image Processing (ICIP)}.\hskip 1em plus 0.5em minus 0.4em\relax IEEE,
  2019, pp. 3362--3366.

\bibitem{balakrishnan2013detecting}
G.~Balakrishnan, F.~Durand, and J.~Guttag, ``Detecting pulse from head motions
  in video,'' in \emph{Proceedings of the IEEE conference on computer vision
  and pattern recognition}, 2013, pp. 3430--3437.

\bibitem{icaart}
A.~Subramaniam and K.~Rajitha, ``Estimation of the cardiac pulse from facial
  video in realistic conditions,'' \emph{ICAART}, pp. 145--153, 2019.

\bibitem{8798700}
S.~Wen, H.~Wei, Z.~Yan, Z.~Guo, Y.~Yang, T.~Huang, and Y.~Chen,
  ``Memristor-based design of sparse compact convolutional neural network,''
  \emph{IEEE Transactions on Network Science and Engineering}, vol.~7, no.~3,
  pp. 1431--1440, 2020.

\bibitem{8824944}
G.~Yuan, X.~Ma, C.~Ding, S.~Lin, T.~Zhang, Z.~S. Jalali, Y.~Zhao, L.~Jiang,
  S.~Soundarajan, and Y.~Wang, ``An ultra-efficient memristor-based dnn
  framework with structured weight pruning and quantization using admm,'' in
  \emph{2019 IEEE/ACM International Symposium on Low Power Electronics and
  Design (ISLPED)}, 2019, pp. 1--6.

\bibitem{lecun1990optimal}
Y.~LeCun, J.~S. Denker, and S.~A. Solla, ``Optimal brain damage,'' in
  \emph{Advances in neural information processing systems}, 1990, pp. 598--605.

\bibitem{hassibi1993optimal}
B.~Hassibi, D.~G. Stork, and G.~J. Wolff, ``Optimal brain surgeon and general
  network pruning,'' in \emph{IEEE international conference on neural
  networks}.\hskip 1em plus 0.5em minus 0.4em\relax IEEE, 1993, pp. 293--299.

\bibitem{lee2018snip}
N.~Lee, T.~Ajanthan, and P.~H. Torr, ``Snip: Single-shot network pruning based
  on connection sensitivity,'' \emph{arXiv preprint arXiv:1810.02340}, 2018.

\bibitem{subramaniam2020n2nskip}
A.~Subramaniam and A.~Sharma, ``N2nskip: Learning highly sparse networks using
  neuron-to-neuron skip connections.'' in \emph{BMVC}, 2020.

\bibitem{memNeuro1}
W.~Huh, D.~Lee, and C.-H. Lee, ``Memristors based on 2d materials as an
  artificial synapse for neuromorphic electronics,'' \emph{Advanced Materials},
  vol.~32, no.~51, p. 2002092, 2020.

\bibitem{memNeuro2}
Y.~Li, Z.~Wang, R.~Midya, Q.~Xia, and J.~J. Yang, ``Review of memristor devices
  in neuromorphic computing: materials sciences and device challenges,''
  \emph{Journal of Physics D: Applied Physics}, vol.~51, no.~50, p. 503002,
  2018.

\bibitem{memNeuro3}
L.~Xia, B.~Li, T.~Tang, P.~Gu, P.-Y. Chen, S.~Yu, Y.~Cao, Y.~Wang, Y.~Xie, and
  H.~Yang, ``Mnsim: Simulation platform for memristor-based neuromorphic
  computing system,'' \emph{IEEE Transactions on Computer-Aided Design of
  Integrated Circuits and Systems}, vol.~37, no.~5, pp. 1009--1022, 2017.

\bibitem{memNeuro4}
S.~H. Jo, T.~Chang, I.~Ebong, B.~B. Bhadviya, P.~Mazumder, and W.~Lu,
  ``Nanoscale memristor device as synapse in neuromorphic systems,'' \emph{Nano
  letters}, vol.~10, no.~4, pp. 1297--1301, 2010.

\bibitem{memNeuro5}
M.~Hu, H.~Li, Y.~Chen, Q.~Wu, G.~S. Rose, and R.~W. Linderman, ``Memristor
  crossbar-based neuromorphic computing system: A case study,'' \emph{IEEE
  transactions on neural networks and learning systems}, vol.~25, no.~10, pp.
  1864--1878, 2014.

\bibitem{raj2019programming}
P.~M.~P. Raj, A.~Subramaniam, S.~Priya, S.~Banerjee, and S.~Kundu,
  ``Programming of memristive artificial synaptic crossbar network using pwm
  techniques,'' \emph{Journal of Circuits, Systems and Computers}, vol.~28,
  no.~12, p. 1950201, 2019.

\bibitem{boppidi2020implementation}
P.~K.~R. Boppidi, V.~J. Louis, A.~Subramaniam, R.~K. Tripathy, S.~Banerjee, and
  S.~Kundu, ``Implementation of fast ica using memristor crossbar arrays for
  blind image source separations,'' \emph{IET Circuits, Devices \& Systems},
  vol.~14, no.~4, pp. 484--489, 2020.

\end{thebibliography}

% \bibitem{b1} G. Eason, B. Noble, and I. N. Sneddon, ``On certain integrals of Lipschitz-Hankel type involving products of Bessel functions,'' Phil. Trans. Roy. Soc. London, vol. A247, pp. 529--551, April 1955.
% \bibitem{b2} J. Clerk Maxwell, A Treatise on Electricity and Magnetism, 3rd ed., vol. 2. Oxford: Clarendon, 1892, pp.68--73.
% \bibitem{b3} I. S. Jacobs and C. P. Bean, ``Fine particles, thin films and exchange anisotropy,'' in Magnetism, vol. III, G. T. Rado and H. Suhl, Eds. New York: Academic, 1963, pp. 271--350.
% \bibitem{b4} K. Elissa, ``Title of paper if known,'' unpublished.
% \bibitem{b5} R. Nicole, ``Title of paper with only first word capitalized,'' J. Name Stand. Abbrev., in press.
% \bibitem{b6} Y. Yorozu, M. Hirano, K. Oka, and Y. Tagawa, ``Electron spectroscopy studies on magneto-optical media and plastic substrate interface,'' IEEE Transl. J. Magn. Japan, vol. 2, pp. 740--741, August 1987 [Digests 9th Annual Conf. Magnetics Japan, p. 301, 1982].
% \bibitem{b7} M. Young, The Technical Writer's Handbook. Mill Valley, CA: University Science, 1989.
% \end{thebibliography}
% \vspace{12pt}

\end{document}